\documentclass[%
prl,
 reprint,
superscriptaddress,
 amsmath,amssymb,
 aps,
]{revtex4-2}

\usepackage[dvipdfmx]{graphicx}
\usepackage{dcolumn}% Align table columns on decimal point
\usepackage{bm}% bold math
\usepackage{xcolor}
\usepackage[colorlinks=true,citecolor=red,linkcolor=blue,urlcolor=blue]{hyperref}

\usepackage{physics}

\DeclareMathOperator{\Div}{div}

\begin{document}

%\preprint{APS/123-QED}

\title{Lorentz Reciprocal Theorem in Fluids with Odd Viscosity}

\author{Yuto Hosaka}
\email{yuto.hosaka@ds.mpg.de}
\affiliation{Max Planck Institute for Dynamics and Self-Organization (MPI-DS), Am Fassberg 17, 37077 G\"{o}ttingen, Germany}

\author{Ramin Golestanian} 
\email{ramin.golestanian@ds.mpg.de}
\affiliation{Max Planck Institute for Dynamics and Self-Organization (MPI-DS), Am Fassberg 17, 37077 G\"{o}ttingen, Germany}
\affiliation{Rudolf Peierls Centre for Theoretical Physics, University of Oxford, Oxford OX1 3PU, UK}
\affiliation{Institute for the Dynamics of Complex Systems, University of G\"{o}ttingen, 37077 G\"{o}ttingen, Germany}

\author{Andrej Vilfan} 
\email{andrej.vilfan@ds.mpg.de}
\affiliation{Max Planck Institute for Dynamics and Self-Organization (MPI-DS), Am Fassberg 17, 37077 G\"{o}ttingen, Germany}
\affiliation{Jo\v{z}ef Stefan Institute, 1000 Ljubljana, Slovenia}

\begin{abstract}
The Lorentz reciprocal theorem---that is used to study various transport phenomena in hydrodynamics---is violated in chiral active fluids that feature odd viscosity with broken time-reversal and parity symmetries. Here we show that the theorem can be generalized to fluids with odd viscosity by choosing an auxiliary problem with the opposite sign of the odd viscosity. We demonstrate the application of the theorem to two categories of microswimmers. Swimmers with prescribed surface velocity are not affected by odd viscosity, while those with prescribed active forces are. In particular, a torque-dipole can lead to directed motion. % 615 characters
\end{abstract}

\maketitle

\begin{figure}[b]
    \includegraphics[width=\columnwidth]{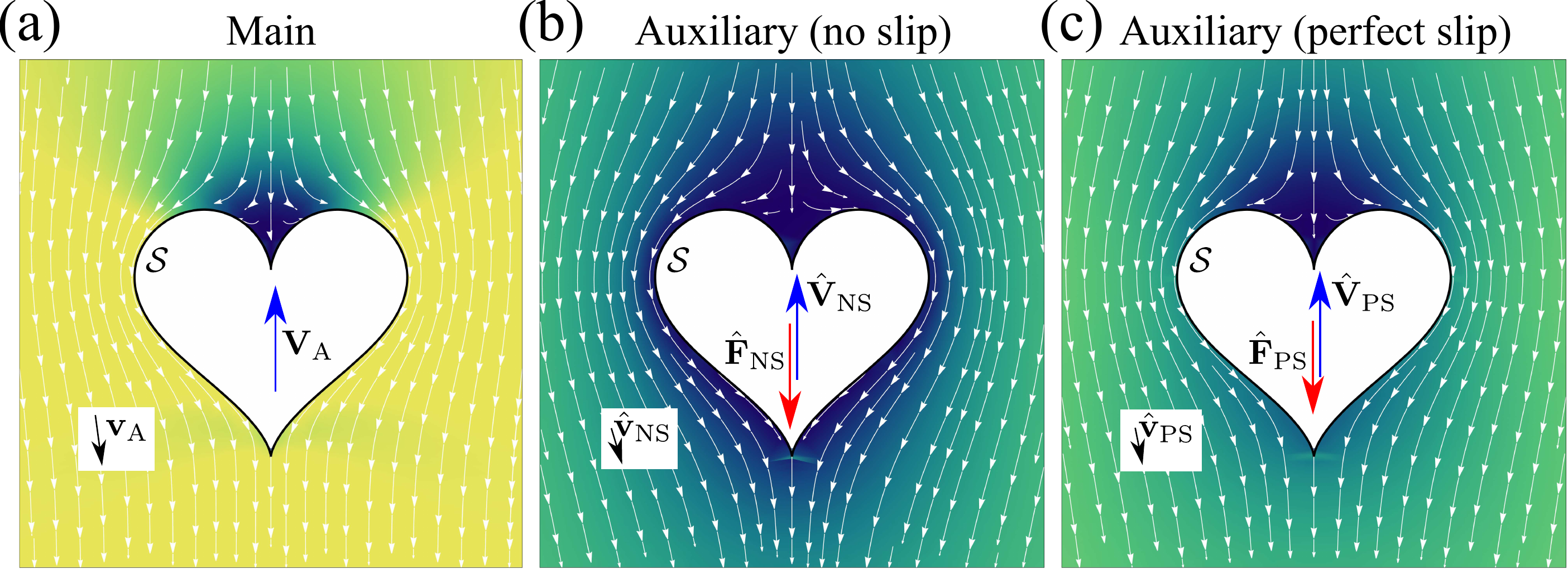}
    \caption{
    The Lorentz reciprocal theorem links the main problem (a) of a body moving with velocity $\mathbf{V}_{\rm A}$, fluid velocity $\mathbf{v}_{\rm A}$ in the co-moving frame, and fluid viscosity $\boldsymbol{\eta}$ with an auxiliary problem (b,c) of a body with the same shape and the velocity $\hat{\mathbf{V}}$, fluid velocity $\hat{\mathbf{v}}$, and viscosity $\hat{\boldsymbol{\eta}}$. Here, the main problem represents an active swimmer and the auxiliary problems in panels (b) and (c) represent bodies with no-slip (NS) and perfect-slip (PS) boundary conditions, respectively.
    \label{fig:Fig1}
}\end{figure}

The Lorentz reciprocal theorem is a powerful and versatile principle in low Reynolds number fluid dynamics~\cite{lorentz1896general, masoud2019}. It provides an integral identity that connects a main problem -- typically the problem to be solved -- to an auxiliary problem -- typically one with a known solution (see Fig.~\ref{fig:Fig1}). The reciprocal theorem allows one to directly determine an integral quantity in the main problem without explicitly solving the boundary-value problem. For instance, one can determine the self-propulsion speed of a microswimmer with the knowledge of its surface velocity~\cite{stone1996propulsion,Najafi2005} -- a common way to solve the problem of phoretic active particles \cite{golestanian2007designing, brady2011particle, mozaffari2016self, nasouri2020exact}, where it is even possible to completely bypass the calculation of the slip velocity~\cite{lammert2016bypassing} or to consider non-axially-symmetric shapes \cite{Poehnl2021}. Other applications include the derivation of boundary integral equations~\cite{pozrikidis1992}, perturbative solutions of flows around bodies that deviate little from exactly solvable ones, flows in porous media or with a slip boundary condition on the surface~\cite{masoud2019}, the presence of elastic surfaces~\cite{DaddiMoussaIder.Stone2018}, the derivation of a lower bound on the dissipation by microswimmers~\cite{Nasouri.Golestanian2021}, and so on. A direct consequence of the reciprocal theorem is the reciprocal response of a system of particles immersed in a fluid to the forces acting on them, which manifests itself in the symmetry of the many-body mobility tensor (or its inverse, the resistance)~\cite{happel2012low}. The latter can be, however, also interpreted as a fundamental property of systems with time-reversal symmetry, related to the Onsager reciprocity~\cite{doi2013soft, doi2015onsager}.

The reciprocal theorem has also been generalized beyond the low-Reynolds-number limit and applied to inertial~\cite{leal1980particle, masoud2019}, compressible~\cite{cunha2003mathematical}, micropolar~\cite{brenner1996lorentz}, and complex fluids~\cite{xu2019generalized}.
Since it is closely related to Green's second identity in vector calculus, it also has a plethora of counterparts in other classical field theories including aerodynamics, acoustics, convective heat and mass transfer, continuum mechanics, electrostatics and electromagnetic waves~\cite{masoud2019}. In electrodynamics, materials with broken Lorentz reciprocity have also been widely studied~\cite{caloz2018electromagnetic}. It has been suggested that broken reciprocity can be used to overcome the time-bandwidth limits on the performance of a device~\cite{tsakmakidis2017breaking}.

In chiral active fluids, the breaking of the time-reversal and parity symmetries results in a peculiar transport coefficient called \textit{odd viscosity}~\cite{avron1995, avron1998, banerjee2017}.
The viscosity, which does not contribute to the fluid energy dissipation, leads to novel dynamics~\cite{hosaka2022nonequilibrium, fruchart2023odd}, such as nonreciprocal (transverse) transports~\cite{lapa2014, hosaka2021nonreciprocal, hosaka2021hydrodynamic, khain2022, lou2022odd, hosaka2023pair, yuan2023stokesian, lier2023lift}, free-surface dynamics~\cite{ganeshan2017, abanov2020hydrodynamics, jia2022incompressible}, or chiral edge currents characterized by topological protection~\cite{souslov2019}, akin to quantum Hall systems.
Chirality is prevalent across various scales in nonequilibrium systems, allowing odd viscosity to occur in systems ranging from electron fluids~\cite{berdyugin2019measuring} to biological~\cite{yamauchi2020, markovich2021} and geophysical flows~\cite{delplace2017topological, tauber2019}.
Indeed, this viscosity has been identified in fluids of spinning particles~\cite{soni2019odd, hargus2020time, han2021fluctuating}, in chiral suspensions~\cite{zhao2022odd}, or even in tabletop experiments at macroscopic scales~\cite{yang2021topologically}.
The lack of time-reversal symmetry leads to an asymmetric response and thus to an asymmetric mobility tensor~\cite{hosaka2021hydrodynamic, yuan2023stokesian, lier2023lift}.
Several works have explicitly shown that the Lorentz reciprocal theorem in its classical form is violated in fluids with odd viscosity~\cite{hosaka2021nonreciprocal, khain2022, fruchart2023odd, yuan2023stokesian}.
A generalization that has been proposed for odd-viscous fluids splits the stress tensor in the auxiliary problem into two parts and reverses the sign in the odd part~\cite{hosaka2021nonreciprocal, yuan2023stokesian}.
The limitation of this approach is that the odd viscosity leads to an additional volume integral in the theorem, which not only breaks the symmetric form between the main and auxiliary problems, but also defeats the dimension-reduction nature of the reciprocal theorem and its power to derive integral quantities in one of the problems entirely without solving the flow equations~\cite{masoud2019}.

In this Letter, we show that the Lorentz reciprocal theorem can be generalized to hold in the presence of odd viscosity. To demonstrate its applicability, we use it to determine the swimming velocity of two categories of microswimmers in a Stokes fluid with odd viscosity: first to swimmers with an imposed slip velocity on their surface and then to swimmers with a prescribed tangential force density. We solve both problems for spherical swimmers in a 3D fluid as well as disk-shaped swimmers in 2D. In both cases, we show that odd viscosity does not affect the motion of swimmers with prescribed surface velocity, but it does affect those with prescribed propulsive forces. A particularly interesting case is a ``twister'' which propels itself by exerting only a torque dipole on the fluid.

\textit{Generalized reciprocal theorem.}---We begin our derivation by first showing that the Lorentz reciprocal theorem can be extended to hold in fluids with generalized viscosity tensors.
Consider two force-free Stokes flows: the main flow described by $\Div\boldsymbol{\sigma}=\mathbf{0}$ and $\nabla\cdot\mathbf{v}=0$ and the auxiliary flow (denoted by~$\hat{}\,$) described by $\Div\hat{\boldsymbol{\sigma}}=\mathbf{0}$ and $\nabla\cdot\hat{\mathbf{v}}=0$, where the divergence is defined as $(\Div\boldsymbol{\sigma})_i\equiv\partial_j\sigma_{ij}$.
Here, $\sigma_{ij}=-p\delta_{ij}+\eta_{ijk\ell}\partial_\ell v_k$ and $\hat{\sigma}_{ij}=-\hat{p}\delta_{ij}+\hat{\eta}_{ijk\ell}\partial_\ell\hat{v}_k$ are the stress fields, $\mathbf{v}$ is the velocity field, $p$ is the pressure, $\boldsymbol{\eta}$ is the fluid viscosity tensor.
Note that, unlike in the regular derivation, we allow different viscosities between the main and auxiliary problems. 
We start with the identity $\partial_j (\hat{\sigma}_{ij}v_i)=(\partial_j\hat{\sigma}_{ij})v_i+ \hat{\sigma}_{ij}\partial_jv_i$ where the first term vanishes because $\Div\hat{\boldsymbol{\sigma}}=\mathbf{0}$ and the second can be written as $-\hat p \delta_{ij} \partial_j v_i+\hat{\eta}_{k\ell ij}(\partial_\ell v_k)(\partial_j\hat{v}_i)$. 
In this expression, the first term vanishes because the flow is divergence-free, $\nabla\cdot\mathbf{v}=0$, leaving $\partial_j (\hat{\sigma}_{ij}v_i)=\hat{\eta}_{k\ell ij}(\partial_\ell v_k)(\partial_j\hat{v}_i)$. By subtracting a similar expression in which the main and the auxiliary problems are swapped, we obtain the identity 
\begin{align}
    \partial_j (v_i\hat{\sigma}_{ij})
    -
    \partial_j (\hat{v}_i\sigma_{ij}) 
    =
    (\hat{\eta}_{k\ell ij}-\eta_{ijk\ell})
    (\partial_\ell v_k)(\partial_j\hat{v}_i).
    \label{eq:LRTright}
\end{align}
In a standard isotropic fluid, the viscosity has the form $\eta_{ijk\ell}=\eta (\delta_{ik}\delta_{j\ell}+\delta_{i\ell}\delta_{jk})$ and by choosing equal viscosities in the main and the auxiliary problems, $\eta=\hat \eta$, the RHS of Eq.~(\ref{eq:LRTright}) vanishes.
Integrating the LHS over the fluid volume leads to the venerable Lorentz reciprocal theorem~\cite{happel2012low, masoud2019}.

In the following we show that the Lorentz reciprocal theorem with any viscosity tensor can be rescued with the right choice of the viscosity tensor in the auxiliary problem, namely such that 
\begin{align}
\label{eq:etasymmetry}
\hat{\eta}_{k\ell ij}=\eta_{ijk\ell}.
\end{align}Then the RHS of Eq.~(\ref{eq:LRTright}) is zero as in the classical case. By integrating the identity over the fluid volume and using the divergence theorem to obtain corresponding surface integrals over all bounding surfaces $\mathcal{S}$~\cite{masoud2019}, we recover the Lorentz reciprocal theorem
\begin{align}
    \int_\mathcal{S} dS\,
    \mathbf{v}\cdot\hat{\boldsymbol{\sigma}}\cdot\mathbf{n}
    =
    \int_\mathcal{S} dS\,
    \hat{\mathbf{v}}\cdot\boldsymbol{\sigma}\cdot\mathbf{n},
    \label{eq:LRT}
\end{align}
where $\mathbf{n}$ is a surface normal pointing into the fluid and $\mathbf{v}\cdot\hat{\boldsymbol{\sigma}}\cdot\mathbf{n}=v_i \hat\sigma_{ij} n_j$.

For a fluid with odd viscosity, the viscosity tensor $\boldsymbol{\eta}=\boldsymbol{\eta}^{\rm e}+\boldsymbol{\eta}^{\rm o}$
consists of a symmetric (even) part $\eta^{\rm e}_{ijk\ell}=\eta^{\rm e}_{k\ell ij}$ and an antisymmetric (odd) part $\eta^{\rm o}_{ijk\ell}=-\eta^{\rm o}_{k\ell ij}$ \cite{avron1995, avron1998}. 
Condition (\ref{eq:etasymmetry}) is therefore fulfilled if the even component in the auxiliary problem is the same as in the main problem, whereas the odd component changes sign
\begin{align}
    \hat{\boldsymbol{\eta}}^{\rm e}={\boldsymbol{\eta}}^{\rm e}, \quad
    \hat{\boldsymbol{\eta}}^{\rm o}=-{\boldsymbol{\eta}}^{\rm o}.
    \label{eq:condition}
\end{align}
These relations state that one has to reverse the sign of the odd viscosity in the auxiliary problem to obtain the quantities in the main problem.
Note that the reciprocal theorem also holds in a fluid with rotational viscosity where a torque appears in response to the vorticity~\cite{epstein2020}, allowing for a local relaxation dynamics of the angular momentum~\cite{de2013non, fruchart2023odd}. 
The viscosity tensor then takes the form $\eta^{\rm R}_{ijk\ell}=\eta^{\rm R}(\delta_{ik}\delta_{j\ell}-\delta_{i\ell}\delta_{jk})$, which gives $\hat{\eta}^{\rm R}=\eta^{\rm R}$, consistent with the dissipative nature of rotational viscosity. The theorem can also be extended to account for body forces acting on the fluid and for multiphase or compressible fluids~\cite{SM}\nocite{hickey2021ciliary}.

%%%%%%%%%%%%%%%%%%%%%%%%%%%%%%%%%%%%%%%%%
\textit{Surface-driven microswimmers.}---As a showcase application of the generalized reciprocal theorem, we now use it to determine the velocity of a microswimmer in a fluid with odd viscosity (see Fig.~\ref{fig:Fig1}). 
We choose the main problem to represent the self-propelled swimmer with velocity $\mathbf{V}_{\rm A}$ and the auxiliary problem as a passive body of the same shape dragged with velocity $\hat{\mathbf{V}}$. We apply the reciprocal theorem [Eq.~(\ref{eq:LRT})] to the fluid surrounding the swimmer. Since both velocities vanish at large distances, the  integrals are reduced to the surface of the swimmer:
\begin{align}
    \int_\mathcal{S} dS\, (\mathbf{V}_{\rm A}+\mathbf{v}_{\rm A})\cdot\hat{\mathbf{f}}
    =
    \int_\mathcal{S} dS\, 
    (\hat{\mathbf{V}}+\hat{\mathbf{v}})\cdot\mathbf{f}_{\rm A},
    \label{eq:LRTswimm}
\end{align}
where $\hat{\mathbf{f}}=\hat{\boldsymbol{\sigma}}\cdot\mathbf{n}$ is the traction, defined as the force density exerted by the fluid on the body, and $\mathbf{v}_{\rm A}$ and $\hat{\mathbf{v}}$ are the fluid velocities in the co-moving frame for the main and the auxiliary problems, respectively.
From the force-free condition of the swimmer ($\mathbf{F}_{\rm A}=\mathbf{0}$), we find $\int_\mathcal{S} dS\, 
\hat{\mathbf{V}}\cdot\mathbf{f}_{\rm A}=\hat{\mathbf{V}}\cdot\mathbf{F}_{\rm A}=0$ in Eq.~(\ref{eq:LRTswimm}).

We consider two types of surface-driven microswimmers: (i) The commonly studied swimmer models have a prescribed effective slip velocity $\mathbf{v}_{\rm A}$ at their surface, which is not influenced by the actual tractions $\mathbf{f}_{\rm A}$. (ii) In the opposite limit, the active mechanism in the swimmer imposes prescribed tangential tractions $\mathbf{f}^\|_{\rm A}=(\mathbf{I}-\mathbf{nn})\cdot\mathbf{f}_{\rm A}$ on the surface, while the normal components  $f^\perp_{\rm A}=\mathbf{n}\cdot\mathbf{f}_{\rm A}$ adjust themselves to satisfy the condition of vanishing normal velocity at the surface. The choice of the auxiliary problem differs between the two scenarios.

To solve the problem of a swimmer with prescribed surface velocity $\mathbf{v}_{\rm A}$, we use a body with a no-slip (NS) boundary condition as an auxiliary problem.
The condition $\hat{\mathbf{v}}_{\rm NS}=\mathbf{0}$ at the swimmer surface simplifies Eq.~(\ref{eq:LRTswimm}) to~\cite{stone1996propulsion, masoud2019}
\begin{align}
    \mathbf{V}_{\rm A}\cdot\hat{\mathbf{F}}_{\rm NS}
    =
    -\int_\mathcal{S} dS\, \mathbf{v}_{\rm A}\cdot\hat{\mathbf{f}}_{\rm NS}.
    \label{eq:vdriven}
\end{align}

Next, we apply the theorem to  a swimmer that is driven by a prescribed active force density $\mathbf{f}^\|_{\rm A}$ on its surface~\cite{daddi2023minimum}. In this case, we choose the auxiliary problem as a body with a perfect-slip (PS) boundary with zero tangential traction $\hat{\mathbf{f}}_{\rm PS}^\|=\mathbf{0}$ and zero normal velocity $\hat{\mathbf{v}}_{\rm PS} \cdot \mathbf{n}=0$. This type of boundary condition applies, for instance, to an idealized air bubble in a fluid. 
Since $\mathbf{v}_{\rm A}$ is tangential at the surface, 
we have $\mathbf{v}_{\rm A}\cdot\hat{\mathbf{f}}_{\rm PS}=0$ and Eq.~(\ref{eq:LRTswimm}) simplifies to
\begin{align}
    \mathbf{V}_{\rm A}\cdot\hat{\mathbf{F}}_{\rm PS}
    =
    \int_\mathcal{S} dS\, \hat{\mathbf{v}}_{\rm PS}\cdot\mathbf{f}_{\rm A}.
    \label{eq:fdriven}
\end{align}

Equations~(\ref{eq:vdriven}) and (\ref{eq:fdriven}) determine the swimming velocity $\mathbf{V}_{\rm A}$ of the swimmer driven by its surface velocity and active force density, respectively, and these relations also hold for purely 2D fluids without momentum decay~\cite{squires2006breaking, elfring2015note, masoud2019}. In the following, we will apply them to spherical swimmers in 3D and to disk-shaped swimmers in 2D fluids with odd viscosity.

\begin{figure}
    \includegraphics[width=\columnwidth]{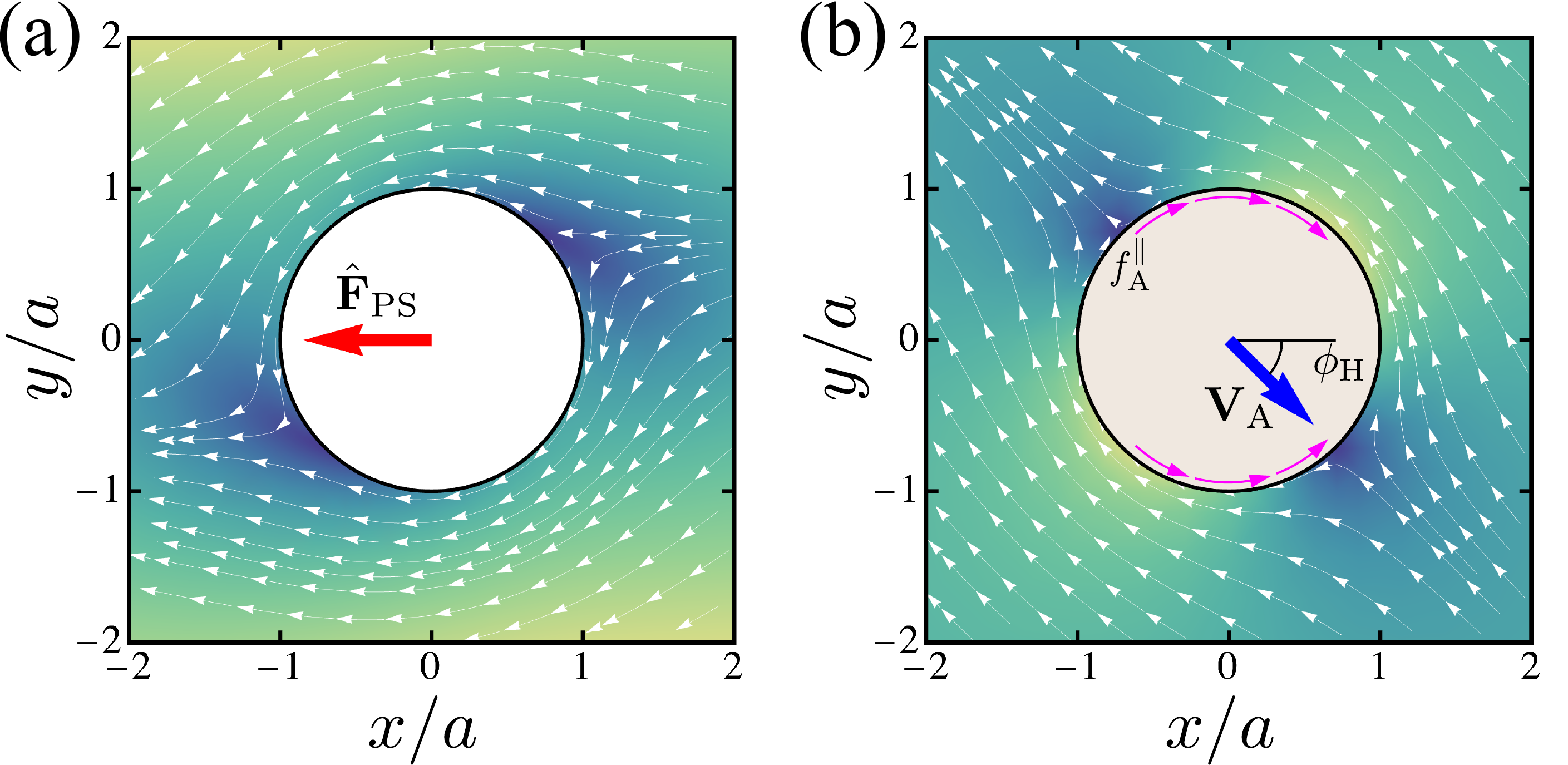}
    \caption{
    The 2D flow fields (a) around a perfect-slip (PS) disk for $\hat{\eta}^{\rm o}=-\hat{\eta}^{\rm e}$, which is used as an auxiliary problem, and (b) generated by a swimmer with prescribed active force density $f_{\rm A}^\|=-f_0\sin\phi$ for $\eta^{\rm o}=\eta^{\rm e}$, moving at an angle $\phi_{\rm H}=-\pi/4$.
    \label{fig:Fig2}
}\end{figure}

%%%%%%%%%%%%%%%%%%%%%
\textit{3D odd flows.}---We now solve the auxiliary passive problems for spherical swimmers in 3D -- first with the no-slip and then with the perfect-slip boundary conditions.
In 3D fluids, multiple independent odd viscosities can arise~\cite{khain2022, fruchart2023odd}.
Here we use a minimal model motivated by a system with aligned spinning microscopic components, which has been shown by microscopic theory to have a single odd viscosity (see Ref.~\cite{markovich2021}).
This system uniquely breaks parity symmetry while maintaining cylindrical symmetry about some axis (hereafter the $z$-axis).  
Then $\boldsymbol{\eta}^{\rm o}$ can be characterized by an odd viscosity $\eta^{\rm o}$, which gives the stress tensor~\cite{markovich2021, oddnotation}
\begin{align}
\sigma_{ij}=-p\delta_{ij}+
2\eta^{\rm e}E_{ij}
+
\eta^{\rm o}
    (
    \epsilon_{zik}E_{kj}
    +
    \epsilon_{zjk}E_{ki}),
    \label{eq:3Dstress}
\end{align}
where $\eta^{\rm e}$ is the even (shear) viscosity, $E_{ij}=(\partial_iv_j+\partial_jv_i)/2$ is the strain-rate tensor, and $\boldsymbol{\epsilon}$ is the 3D Levi-Civita tensor.
Due to its nondissipative nature, $\eta^{\rm o}$ can be either positive or negative depending on the chirality of the fluid, unlike the always positive $\eta^{\rm e}$.
Using $\Div\boldsymbol{\sigma}=\mathbf{0}$ and $\nabla\cdot\mathbf{v}=0$, one obtains the corresponding ``odd" Stokes equation~\cite{markovich2021, khain2022}.
In the limit of $\lambda=\eta^{\rm o}/\eta^{\rm e}\ll1$, the primary Green's function of the 3D odd Stokes equation can be expressed with $\mathbf{G}=\mathbf{G}^{\rm e} + \mathbf{G}^{\rm o}$ where 
$\mathbf{G}^{\rm e}=(\mathbf{I}+\mathbf{rr}/r^2)/r$ and $\mathbf{G}^{\rm o}=- (\lambda/2)\boldsymbol{\epsilon}\cdot(\mathbf{e}_z-z\mathbf{r}/r^2)/r$~\cite{khain2022, yuan2023stokesian, oddnotation}.
When $\lambda\neq0$, the nonreciprocal response $G_{ij}\neq G_{ji}$ arises due to the violation of the Lorentz reciprocity.

The flow around a no-slip sphere with radius $a$, fixed at $\mathbf{r}=\mathbf{0}$, can be represented to the first order in $\hat{\lambda}=\hat{\eta}^{\rm o}/\hat{\eta}^{\rm e}$ as a superposition of a Stokeslet and a source dipole  (see Supplemental Material for the derivation \cite{SM})
\begin{align}
    \hat{\mathbf{v}}_{\rm NS}
    =&
    -
    \hat{\mathbf{V}}
    +
    \frac{3a}{4}
    \left(
    \mathbf{G}
    +
        \frac{a^2}{3}
    \mathbf{D}
    \right)
    \cdot\hat{\mathbf{V}}\nonumber\\
    &-
    \hat{\lambda}\frac{3a}{16}
    \left(
    \mathbf{G}^{\rm e}
    +
        \frac{a^2}{3}
    \mathbf{D}^{\rm e}
    \right)
    \cdot
    (\mathbf{e}_z\times\hat{\mathbf{V}})
    ,
    \label{eq:vns}
\end{align}
with $\mathbf{D}=\frac{1}{2}\nabla^2\mathbf{G}$ and $\mathbf{D}^{\rm e}=\frac{1}{2}\nabla^2\mathbf{G}^{\rm e}$.
In the above, $-\hat{\mathbf{V}}$ is the uniform velocity field for $r\to\infty$. 
Inserting Eq.~(\ref{eq:vns}) and the corresponding pressure~\cite{SM} into Eq.~(\ref{eq:3Dstress}) yields the force acting on the particle
\begin{align}
\hat{\mathbf{F}}_{\rm NS} 
= -6\pi\hat{\eta}^{\rm e}a\bigg(\mathbf{I}+\frac{\hat{\lambda}}{4}\boldsymbol{\epsilon}\cdot\mathbf{e}_z\bigg)\cdot\hat{\mathbf{V}}
= 4\pi a^2 \hat{\mathbf{f}}_{\rm NS}.
\label{eq:fNS3D}
\end{align}
The force is modified by $\hat{\eta}^{\rm o}$ when the sphere moves laterally in the $xy$-plane, while the motion along the $z$-axis causes no corrections, resulting in Stokes' law~\cite{happel2012low, kim2013microhydrodynamics}.

The second auxiliary problem we need to solve is the flow around a spherical body with the perfect-slip boundary. 
The flow can be expressed as~\cite{SM}
\begin{align}
    \hat{\mathbf{v}}_{\rm PS}
    &=
    -\hat{\mathbf{V}}
    +
    \frac{a}{2}
    \mathbf{G}
    \cdot
    \hat{\mathbf{V}}
    +
        \frac{a^3}{6}
    \left[
    \hat{\lambda}
    \mathbf{D}^{\rm e}\cdot
    (\mathbf{e}_z\times\hat{\mathbf{V}})
    -
    \mathbf{D}^{\rm o}\cdot
    \hat{\mathbf{V}}
    \right]
    \label{eq:vps},
\end{align}
with $\mathbf{D}^{\rm o}=\frac{1}{2}\nabla^2\mathbf{G}^{\rm o}$.
For conventional fluids without $\hat{\eta}^{\rm o}$, the perfect-slip boundary conditions are satisfied solely with a simple Stokeslet.
Only the tangential component of Eq.~(\ref{eq:vps}) remains at the surface, which is evaluated as
\begin{align}
    \hat{\mathbf{v}}_{\rm PS}^\| =
    \frac{1}{2}
    \left[
    (\mathbf{nn}-\mathbf{I})\cdot
    (\hat{\mathbf{V}}-\hat{\lambda}\mathbf{e}_z\times\hat{\mathbf{V}})
    -
    \hat{\lambda}\mathbf{e}_z\cdot\mathbf{n}
    \mathbf{n}\times\hat{\mathbf{V}}
    \right]
    .
    \label{eq:3DPSv}
\end{align}
The force acting on the perfect-slip sphere follows as
\begin{align}
    \hat{\mathbf{F}}_{\rm PS}
    =
    -4 \pi\hat{\eta}^{\rm e} a \hat{\mathbf{V}},
    \label{eq:3DPSF}
\end{align}
with no modifications in the leading order of $\hat{\eta}^{\rm o}$, which is consistent with a spherical bubble in a classical fluid~\cite{happel2012low}.

%%%%%%%%%%%%%%%%%%%%%%%%%%%%%%%%%
\textit{2D odd flows.}---As in the 3D case, we first analyze the odd Stokes flow for the no-slip boundary problem and then proceed to discuss the perfect-slip one. 
Due to the Stokes paradox, the velocity around a 2D object diverges at infinity, and the flow problem is therefore ill-posed in the laboratory frame~\cite{happel2012low}.
However, to obtain the swimming velocity of the active swimmer, it is sufficient to establish the relationship between the force $\hat{\mathbf{F}}$ acting on the passive object and its surface velocity $\hat{\mathbf{v}}$ in the co-moving frame, rather than deriving these expressions independently.
In this way, the Stokes paradox is bypassed. 
Note that the solutions of 2D odd flows we will use below are exact for any value of the odd viscosity, unlike 3D odd flows where $\hat{\lambda}\ll1$ is assumed.

The flow around a no-slip disk can be expressed as~\cite{SM}
\begin{align}
    \hat{\mathbf{v}}_{\rm NS} &= 
    \frac{1}{8\pi\hat{\eta}^{\rm e}}\hat{\mathbf{F}}_{\rm NS}
    -
    \frac{1}{4\pi\hat{\eta}^{\rm e}}
    \left(
        \mathbf{g} + \frac{a^2}{2}\mathbf{d}
    \right)
    \cdot\hat{\mathbf{F}}_{\rm NS},
    \label{eq:NS}
\end{align}
where $\mathbf{g}=-\ln(\rho/a)\mathbf{I}+\boldsymbol{\rho}\boldsymbol{\rho}/\rho^2$ and $\mathbf{d}=\frac{1}{2}\nabla^2\mathbf{g}$ are the 2D Stokeslet and source dipole~\cite{pozrikidis1992}, respectively, with $\boldsymbol{\rho}=(x,y)$ and $\rho=|\boldsymbol{\rho}|$.
The flow around a no-slip body is not affected by the odd viscosity, which is a general feature of incompressible fluids in 2D systems~\cite{ganeshan2017}. 
The force is simply found as
\begin{align}
    \hat{\mathbf{F}}_{\rm NS} = 2\pi a\hat{\mathbf{f}}_{\rm NS},
    \label{eq:NSf}
\end{align}
where the traction is constant over the surface, as in the case of the  no-slip sphere in Eq.~(\ref{eq:fNS3D}).

The flow around a circular object with the perfect-slip boundary condition (${\mathbf{n}\cdot\hat{\mathbf{v}}_{\rm PS}=0}$ and $\hat{f}_{\rm PS}^\|=0$ at $\rho=a$) can be exactly expressed as~\cite{SM}
\begin{align}
    \hat{\mathbf{v}}_{\rm PS} 
    =& 
     \frac{1}{4\pi\hat{\eta}^{\rm e}}
     (\mathbf{I}-\mathbf{g})\cdot\hat{\mathbf{F}}_{\rm PS}
     -
    \frac{1}{8\pi\hat{\eta}^{\rm e}} 
    \frac{\hat{\lambda}}{1+\hat{\lambda}^2}\nonumber
     \\
     &
     \times\left(
     \mathbf{I}
     +
     a^2
     \mathbf{d}
     \right)
     \cdot
     \left( 
    \mathbf{e}_z\times\hat{\mathbf{F}}_{\rm PS}
    +
    \hat{\lambda}\hat{\mathbf{F}}_{\rm PS}
    \right)
    .
    \label{eq:PS}
\end{align}
In contrast to the no-slip case, the flow with a perfect-slip boundary condition depends on $\hat{\eta}^{\rm o}$. The surface velocity has only a tangential component, which reads
\begin{align}
    \hat{v}_{\rm PS}^\|
    &=
    \frac{1}{4\pi\hat{\eta}^{\rm e}}
    \frac{1}{{1+\hat{\lambda}^2}}
    (\mathbf{e}_z\times\mathbf{n}
    -
    \hat{\lambda}\mathbf{n})\cdot\hat{\mathbf{F}}_{\rm PS}.
    \label{eq:vpsf}
\end{align}
\begin{figure}
\centering
    \includegraphics[width=\linewidth]{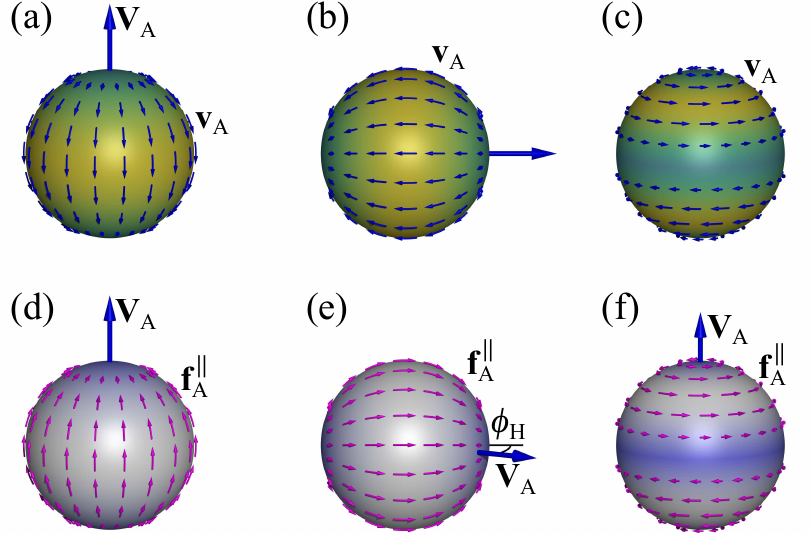}
    \caption{
    Swimming behavior of swimmers with (a,b,c) prescribed velocity $\mathbf{v}_{\rm A}$ and (d,e,f) prescribed traction $\mathbf{f}^\|_{\rm A}$.     The tangential velocity/traction has the meridional (a,d), orthogonal meridional (b,e), or azimuthal (c,f) direction~\cite{SM}. 
    The \textit{twister} (f) moves along the axis defined by odd viscosity by exerting a torque dipole on the fluid.
    \label{fig:Fig3}
}\end{figure}%
Figure~\ref{fig:Fig2}(a) shows the velocity field around a perfect-slip disk [see Eq.~(\ref{eq:PS})].
For $\hat{\eta}^{\rm o}=0$ the resulting streamlines are symmetric with respect to the force direction $\hat{\mathbf{F}}_{\rm PS}$, while for $\hat{\eta}^{\rm o}=-\hat{\eta}^{\rm e}$ the flow develops transverse to $\hat{\mathbf{F}}_{\rm PS}$ and the above symmetry about the $x$-direction breaks down accordingly.
Similarly distorted streamlines have been obtained for an odd-viscous liquid domain moving in a 2D fluid~\cite{hosaka2021hydrodynamic}.

%%%%%%%%%%%%%%%%%%%%%%%%%%%%%%%%%%%%
\textit{Microswimmers' dynamics.}---Having solved the necessary auxiliary problems, now we derive the swimming velocities of spherical and circular microswimmers by employing the Lorentz reciprocal theorem.

For swimmers with a prescribed surface velocity
we use Eqs.~(\ref{eq:vdriven}) and (\ref{eq:fNS3D}), and obtain the swimming velocity as
\begin{align}
    \mathbf{V}_{\rm A} = -\frac{1}{4\pi a^2} \int_\mathcal{S} dS\, \mathbf{v}_{\rm A}.
    \label{eq:v3Drigid}
\end{align}
Similarly,  the 2D swimming velocity follows from Eqs.~(\ref{eq:vdriven}) and (\ref{eq:NSf}) as
\begin{align}
    \mathbf{V}_{\rm A} = -\frac{1}{2\pi a} \int_\mathcal{C} ds\, \mathbf{v}_{\rm A},
    \label{eq:v2Drigid}
\end{align}
where $\mathcal{C}$ denotes the disk perimeter.
Noting that Eqs.~(\ref{eq:v3Drigid}) and (\ref{eq:v2Drigid}) are the same as those for a classical fluid without odd viscosity~\cite{stone1996propulsion, masoud2019, elfring2015note}, we can conclude that velocity-prescribed microswimmers are not affected by odd viscosity [Fig.~\ref{fig:Fig3}(a), (b), and (c)].

In stark contrast to the velocity-prescribed swimmer, the one with prescribed forces is strongly affected by the odd viscosity and it can show intriguing nonreciprocal responses.
Inserting Eqs.~(\ref{eq:3DPSv}) and (\ref{eq:3DPSF}) into Eq.~(\ref{eq:fdriven}) yields the active swimming velocity for $\lambda=\eta^{\rm o}/\eta^{\rm e}\ll1$
\begin{align}
    \mathbf{V}_{\rm A} = 
    \frac{1}{8\pi\eta^{\rm e}a}
    \int_\mathcal{S} dS\,
    \left[
    \mathbf{f}_{\rm A}^\|
    +\lambda
    \mathbf{f}_{\rm A}^\|
    \times
    \left(
    \mathbf{I}
     -
     \mathbf{n}\mathbf{n}
    \right)\cdot\mathbf{e}_z
    \right],
    \label{eq:v3Dbubble}
\end{align}
where we have set $\hat{\eta}^{\rm e}=\eta^{\rm e}$ and $\hat{\eta}^{\rm o}=-\eta^{\rm o}$ or $\hat{\lambda}=-\lambda$.
Expression~(\ref{eq:v3Dbubble}) shows that the positive values of $\lambda$ allow a transverse (Magnus or Hall) transport rotated clockwise by $\pi/2$ from the regular swimming direction.
Analogous to the Hall effect in an electron fluid with a magnetic field~\cite{berdyugin2019measuring}, the longitudinal and transverse velocities give a Hall angle.
For the active force density $\mathbf{f}^\|_{\rm A}=f_0
(\mathbf{I}-\mathbf{n}\mathbf{n})\cdot \mathbf{e}_x$, as in Fig.~\ref{fig:Fig3}(e), the angle is given by $\phi_{\rm H}\approx-\lambda/2$.
Interestingly, vertical motion can also be achieved with purely azimuthal forces.
The lowest mode of such force density, with zero total torque, turns out to be $\mathbf{f}_{\rm A}^\|=f_0\sin(2\theta)\mathbf{e}_\phi$.
We call the swimmer with this profile a \textit{twister} [Fig.~\ref{fig:Fig3}(f)], and it moves vertically with
\begin{align}
    \mathbf{V}_{\rm A} = 
    \lambda\frac{2 a f_0}{15\eta^{\rm e}}
    \mathbf{e}_z.
\end{align}
The nonzero values of $\lambda$ allow for the nonreciprocal response connecting the active torque dipole and its induced motion along the $z$-axis.

For a disk-shaped swimmer in a 2D fluid, we find using Eqs.~(\ref{eq:fdriven}) and (\ref{eq:vpsf}) the swimming velocity as
\begin{align}
    \mathbf{V}_{\rm A} =
    \frac{1}{4\pi\eta^{\rm e}}
    \frac{1}{1+\lambda^2}
    \int_\mathcal{C} ds\,
    \left(
     \mathbf{f}_{\rm A}^\|
     +
     \lambda
     \mathbf{f}_{\rm A}^\|\times
     \mathbf{e}_z
    \right),
    \label{eq:v2Dbubble}
\end{align}
which gives $\phi_{\rm H}=-\arctan(\lambda)$.
The transverse velocity exhibits a nonmonotonic dependence in the odd viscosity and is maximized when $\lambda=1$.
Figure~\ref{fig:Fig2}(b) shows the streamlines induced by a microswimmer with $f_{\rm A}^\|=-f_0\sin\phi$ for $\lambda=1$, moving at an angle $\phi_{\rm H}=-\pi/4$~\cite{SM}.
In 2D odd-viscous fluids, transverse transport has been verified in the context of active rheology~\cite{hosaka2021nonreciprocal, hosaka2021hydrodynamic, reichhardt2022, lier2023lift} or a biased-dipolar microswimmer in a compressible fluid layer~\cite{hosaka2023hydrodynamics}.

%%%%%%%%%%%%%%%%%%%%%%%%%%%%%%%%%%%%%%%%%%%%%%%%%%%%
In conclusion, we have derived a generalized form of the Lorentz reciprocal theorem for fluids with odd viscosity and demonstrated its application to several categories of microswimmers. Our work facilitates the solution of a number of swimming and flow problems in fluids with odd viscosity and should be applicable to various chiral active systems, such as spinning colloidal suspensions~\cite{soni2019odd}, chiral living assemblies~\cite{drescher2009dancing, bililign2022motile, tan2022odd} or intracellular or multicellular environments~\cite{chen2020motor, yamauchi2020}, as well as ferrofluids~\cite{reynolds2023} or electron fluids~\cite{avron1995, berdyugin2019measuring}. In view of the number of related reciprocal theorems in other fields,  generalizations to other systems with broken time-reversal symmetries remain an intriguing question for the future.

We acknowledge support from the Max Planck Society, the Max Planck School Matter to Life, and the MaxSynBio Consortium, which are jointly funded by the Federal Ministry of Education and Research (BMBF) of Germany. A.V.\ acknowledges support from the Slovenian Research Agency (Grant No.\ P1-0099).

\bibliography{myref}

\end{document}